\documentstyle[psfig]{mn}

\title[Absorption systems in the 10k 2QZ catalogue]{The 2dF QSO Redshift Survey - VIII. Absorption systems in the 10k catalogue}
\author[P.~J.~Outram et al.]
{P.~J.~Outram$^1$, R.~J.~Smith$^{2,3}$, T.~Shanks$^1$, B.~J.~Boyle$^4$, S.~M.~Croom$^4$,
\newauthor N.~S.~Loaring$^5$,  L.~Miller$^5$ \\
$^1$ Dept. of Physics, University of Durham, South Road, Durham DH1 3LE, UK.\\
$^2$ Astrophysics Research Institute, Liverpool John Moores University, 12 Quays House, Egerton Wharf, Birkenhead, CH41 1LD, UK.\\
$^3$ Research School of Astronomy \& Astrophysics, Mount Stromlo Observatory, Institute of Advanced Studies,\\ Australian National University, Private bag, Weston Creek P.O., ACT 2611, Australia.\\
$^4$ Anglo-Australian Observatory, PO Box 296, Epping, NSW 2121, Australia.\\
$^5$ Dept. of Physics, University of Oxford, Nuclear \& Astrophysics Laboratory, Keble Road, Oxford, OX1 3RH, UK. \\
}

\begin{document}
\maketitle
\begin{abstract} 
We examine the highest S/N spectra from the 2QZ 10k release and identify over 100 new low-ionisation heavy element absorbers; DLA candidates suitable for higher resolution follow-up observations.  These absorption
systems  map the
spatial distribution of high-$z$ metals {\it in exactly  the same volumes} that
the foreground 2QZ QSOs themselves sample and hence the 2QZ gives us the unique
opportunity to directly compare the two tracers of large scale structure. We examine the cross-correlation of the two populations to see how they are relatively clustered, and, by considering the colour of the QSOs, detect a small amount of dust in these metal systems.

\end{abstract}

\section{Introduction}

Progress in the study of QSO absorption line systems has gone hand in hand
with the advancement of technology over the past few years. During the
1980's developments in echelle spectroscopy with sensitive electronic
detectors increased the attainable resolution by a factor of ten. The
detailed analysis of the absorption systems in these spectra was
possible through absorption line profile fitting. The increase in
spectral dispersion coupled with the need to obtain a respectable S/N,
however, forced astronomers to observe the same object for many nights
on 4$\,$m-class telescopes, restricting the number of objects it was
possible to observe. In the 1990's the 10$\,$m Keck telescope together 
with its powerful instrument
HIRES enabled us to obtain optical spectra of faint, high redshift
QSOs at unprecedented spectral resolution ($\sim 7\,$km$\,$s$^{-1}$)
and a signal-to-noise ratio in excess of a hundred in a single night. Meanwhile, ultra-violet spectroscopy using the Hubble Space Telescope opened a window on QSO absorbers at low redshift. Using these data, very detailed studies of the chemical properties of the largest absorbers, damped Lyman-alpha systems (DLAs), have provided a wealth
of information about the formation of structure. Understanding the chemical evolutionary history of galaxies, seen here in absorption,
is fundamental to the study of galaxy formation. 

The study of DLAs has suffered from the small number of objects, approximately one hundred, that are currently known. Although detailed studies of individual objects are very revealing, HST imaging of QSO fields has revealed a wide range of
luminosities and morphologies for DLA counterparts. Hence to determine the evolution in the absorption properties of a such mixed population of galaxies, much larger samples of QSO absorbers will be required.

With the recent release of the 2dF QSO Redshift Survey (2QZ) 10k Catalogue (Croom et al. 2001a) the number of known QSOs has suddenly and dramatically increased. With the 2QZ rapidly approaching its target of 25000 QSOs, together with Sloan Digital Sky Survey observations (Fan et al. 1999), unprecedented numbers of new QSO spectra will soon be available from which we can identify and study large numbers of new heavy element absorption systems.

As the 2QZ spectra were taken primarily to confirm the identity of QSOs, and determine their redshift, they have a typical signal-to-noise ratio (S/N) $\sim$10, and a resolution of $\sim$8\AA. Although this is not ideal for absorption line analysis, it is possible to identify strong heavy element absorption systems, especially those with distinctive features, such as Mg$\;${\small II} / Fe$\;${\small II} absorption. 

In this letter we examine the highest S/N spectra from the 2QZ 10k release and identify over 100 new heavy element absorbers, suitable for higher resolution follow-up.  These absorption
systems  map the
spatial distribution of high-$z$ gas and metals {\it in exactly  the same volumes} that
the foreground 2QZ QSOs themselves sample and hence the 2QZ gives us the unique
opportunity to directly compare the two tracers of large scale structure. We examine the cross-correlation of the two populations to see how they are relatively clustered. Finally, we consider the colour of the QSOs containing absorption systems, relative to the 2QZ average and detect a small amount of dust in these metal systems.

\section{2QZ Spectra}

The 2QZ
comprises two $5\times75\deg^2$ declination strips, one at the South
Galactic Pole and one in an equatorial region in the North Galactic
Cap. QSOs candidates were selected by ultra-violet excess (UVX) in the $U$-$B$:$B$-$R$
plane, using APM measurements of UK Schmidt Telescope photographic plates (Smith et al. 2001). 

2QZ objects were observed with the 2-degree Field instrument (Lewis, Glazebrook \& Taylor 1998) on the Anglo-Australian Telescope using the low resolution 300B grating, providing a dispersion of 178.8\AA\ mm$^{-1}$ (4.3\AA\ pixel$^{-1}$) and a resolution of $\sim 8.6$\AA\ ($\sim$2 pixels) over the wavelength range 3700 -- 7500\AA. Each field received approximately 1 hour integration, adjusted to match conditions where possible. 

The data were reduced using the pipeline reduction system 2DFDR (Bailey \& Glazebrook 1999). Following bias subtraction, the spectra from each fibre were extracted and wavelength calibrated using a CuAr+He arc. Sky subtraction was done using a mean sky spectrum determined from between six and twenty sky-dedicated fibres in each frame. 

The reduced 2dF spectra  were identified automatically by \textsc{
autoz} (Miller et al., in preparation), which also determines the redshift of the QSOs. These identifications were confirmed by eye.
Approximately 18000 QSO redshifts have been obtained to date,
making this survey already an order of magnitude larger than any
previous QSO survey.

The primary science goal of the 2QZ is to obtain an accurate measure of large-scale structure via the study of QSO clustering. The observations are therefore optimised to maximise the number of QSO identifications. Hence the S/N and resolution, whilst ideal for identifying broad emission features, are generally too low to detect all but the strongest narrow absorption features. Furthermore, the sky-subtraction possible on a multi-fibre instrument such as 2dF is relatively poor, leading to large uncertainties in the zero-level of the spectra (making equivalent width measurements less reliable), and variable sky lines, particularly at the red end of the spectrum. Having said that, the 10000 QSO spectra released so far are still, as we shall demonstrate, a hugely valuable resource for QSO absorption line studies.

\section{Absorption Systems}

To detect absorption systems in the spectra of the 2QZ QSOs released in the 10k catalogue (Croom et al 2001a), QSOs with $z_{abs} > 0.5$ and S/N$>15$ were selected from the catalogue for examination by eye. Systems identified by two or more absorption features are included in this analysis (due to the resolution of the spectra, doublets such as C$\;${\small IV} $\lambda\lambda 1548,1550$ and Mg$\;${\small II} $\lambda\lambda 2796,2803$ were counted as single features). As the primary aim of this letter is to identify heavy element systems that could correspond to DLA candidates, only systems that exhibit low ionisation absorption features are reported here. Several of the spectra examined were BAL QSOs, or exhibited relatively high ionisation $z_{abs} \sim z_{em}$ systems which are excluded from this analysis. Also excluded were several  C$\;${\small IV} systems, often with associated  Si$\;${\small IV} and  H$\;${\small I} absorption. Finally all unconfirmed single absorption features are excluded, including Ly$\alpha$\  absorbers in the forest of high redshift QSOs that do not exhibit metal features, and individual metal features that most likely correspond to C$\;${\small IV} or Mg$\;${\small II} systems.

\begin{figure}
\centerline{\hbox{\psfig{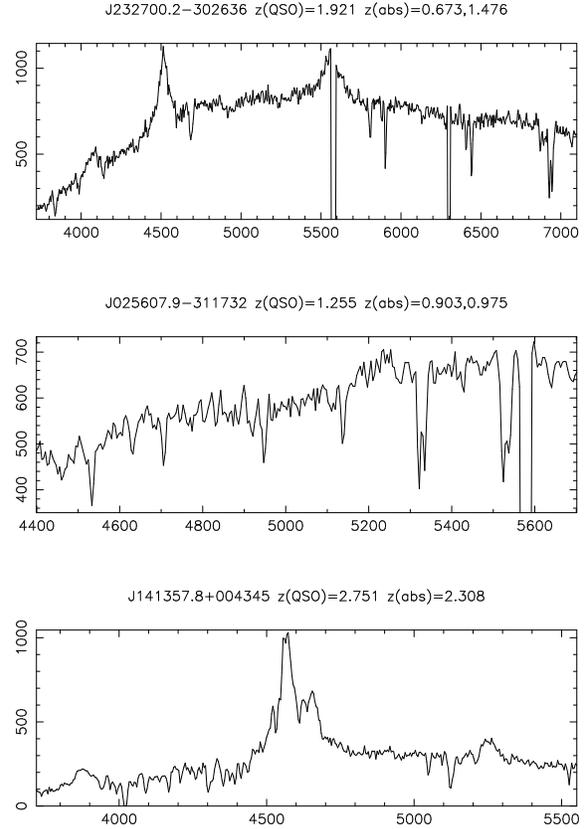}}}
\caption{Example 2QZ spectra containing identified absorption systems. The first spectrum shows the distinctive pattern of Fe$\;$II absorption at 5850\AA\  and 6420\AA, Mg$\;$I at 7060\AA\  and the strong Mg$\;$II doublet at 6930\AA, together with C$\;$IV absorption 3830\AA, Al$\;$II at 4140\AA, and Si$\;$II at 3780\AA\  all due to a system at $z_{abs}=1.476$. A second system lies at lower redshift; the Mg$\;$II doublet at 4680\AA\  is visible, along with Fe$\;$II absorption at 3980\AA, and 4350\AA. Two further examples of Mg/Fe systems can be seen in the second spectrum. The final spectrum has a distinctive Ly$\alpha$\  forest. The H$\;$I $\lambda 1215$ line at 4020\AA\ is a strong DLA candidate with corresponding C$\;$IV absorption at 5120\AA, Al$\;$II at 5525\AA, and Si$\;$II at 5050\AA\ amongst the other ions visible.
}
\label{fig1}
\end{figure}

In total, 137 low-ionisation absorption systems were detected in 1266 QSO spectra. Of these, 120 are Mg$\;${\small II} systems, and the remaining 17 are C$\;${\small IV} systems at higher redshift, which also exhibit low ionisation species, and in 12 cases, strong Ly$\alpha$\  absorption. Of these, two have a rest equivalent width in excess of 10\AA, and hence are strong damped Ly$\alpha$\  line candidates. Three example 2QZ spectra showing absorption systems are shown in Fig.~1, and Table~1 contains a full list of the low ionisation absorption systems. Where possible the rest equivalent width ($W_0$) of the Mg$\;${\small II} $\lambda\lambda 2796,2803$ doublet (\AA), or H$\:${\small I} $\lambda 1215$ for the high redshift systems, is quoted. Due to the low resolution of the data the separate Mg$\;${\small II} features are often not resolved. The $W_0$ determination is not highly reliable due to the relatively poor sky subtraction possible with fibre observations, leading to zero-level uncertainty, particularly at the blue end of the spectrum. Also, there is relatively high continuum uncertainty due to the low S/N and resolution of the data, particularly in the Ly$\alpha$\  forest. Typical uncertainty in the $W_0$ is of order 20\%.

\begin{table}
\caption{Low ionisation 2QZ absorption systems}
\label{list}
\begin{tabular}{lccccl}
\hline  
2QZ QSO ID & $b_J$ & $z_{QSO}$ & $z_{abs}$ &$W_0^2$& Ion$^1$  \\
\hline  
J095605.0-015037 & 19.04 & 1.188 & 1.045 &3.04& efl\\
J095938.2-003501 & 19.42 & 1.872 & 1.598 &4.31& fl\\
J101230.1-010743 & 19.28 & 2.360 & 1.370 &2.83& fl\\
J101556.2-003506 & 18.63 & 2.497 & 1.489 &2.29& cefgl\\
J101636.2-023422 & 19.55 & 1.519 & 0.912 &2.90& fl\\
J101742.3+013216 & 18.35 & 1.457 & 1.313 &1.72$^3$& fl\\
J102645.2-022101 & 19.65 & 2.401 & 1.581 &1.31$^3$& cfgl\\
J103727.9+001819 & 18.48 & 2.466 & 2.301 &14.0$^4$& abcgijkl\\
J105304.0-020114 & 19.96 & 1.527 & 0.888 &5.76& efl\\
J105620.0-000852 & 18.38 & 1.440 & 1.285 &1.70& fl\\
J105811.9-023725 & 18.30 & 2.344 & 2.278 &4.25$^4$& acgikl\\
J110603.4+002207 & 19.06 & 1.659 & 1.018 &2.08& fl\\
J110624.6-004923 & 19.84 & 2.414 & 2.122 &10.3$^4$& acgi\\
J110736.6+000328 & 18.86 & 1.726 & 0.953 &2.65& fl\\
J114101.3+000825 & 19.89 & 1.573 & 0.841 &3.93& fl\\
J115352.0-024609 & 19.42 & 1.835 & 1.204 &2.90& fl\\
J115559.7-015420 & 19.05 & 1.259 & 1.132 &3.75& efl\\
J120455.1+002640 & 19.23 & 1.557 & 0.597 &3.92& efl\\
J120826.9-020531 & 19.35 & 1.724 & 0.761 &5.31& fl\\
J120827.0-014524 & 19.69 & 1.552 & 0.621 &3.13& efl\\
J120836.2-020727 & 18.42 & 1.081 & 0.873 &1.97& efl\\
J120838.1-025712 & 18.70 & 2.327 & 2.314 &2.44$^4$& achijk\\
J121318.9-010203 & 20.14 & 2.515 & 2.415 &5.33$^4$& abcdgijk\\
J121957.7-012615 & 19.30 & 2.636 & 2.562 &3.09$^4$& acgjk\\
J122454.4-012753 & 19.63 & 1.347 & 1.089 &2.14& fl\\
J125031.5+000216 & 19.61 & 2.100 & 1.327 &2.61& fl\\
J125359.6-003227 & 19.48 & 1.689 & 1.503 &5.44& cf\\
J125658.3-002123 & 19.05 & 1.273 & 0.947 &3.62& fl\\
J130019.9+002641 & 18.81 & 1.748 & 1.225 &5.92& fl\\
J130433.0-013916 & 19.56 & 1.596 & 1.410 &8.05& fl\\
J130622.8-014541 & 18.98 & 2.152 & 1.332 &3.67& fl\\
J133052.4+003219 & 18.73 & 1.474 & 1.327 &1.89& fl\\
J134448.0-005257 & 19.52 & 2.083 & 0.932 &5.44& efl\\
J134742.0-005831 & 19.32 & 2.515 & 1.795 &3.69$^5$& bcgikl\\
J135941.1-002016 & 18.57 & 1.389 & 1.120 &2.97& fl\\
J140224.1+003001 & 19.25 & 2.411 & 1.387 &2.85& fl\\
J140710.5-004915 & 19.25 & 1.509 & 1.484 &3.49& fl\\
J141051.2+001546 & 18.94 & 2.598 & 1.170 &4.74& efl\\
J141357.8+004345 & 20.14 & 2.751 & 2.308 &5.81$^4$& abcdgijkl\\
J142847.4-021827 & 18.97 & 1.312 & 1.313 &2.24& fl\\
J144715.4-014836 & 19.27 & 1.606 & 1.354 &2.02& fl\\
\hline
J214726.8-291017 & 19.39 & 1.678 & 0.931 &1.74& fl\\
J214836.0-275854 & 18.40 & 1.998 & 1.112 &1.40& fl\\
J215024.0-282508 & 19.03 & 2.655 & 1.144 &1.88& efl\\
J215034.5-280520 & 18.63 & 1.358 & 1.139 &1.58& fl\\
J215102.9-303642 & 19.78 & 2.525 & 2.488 &7.48$^4$& abcgik\\
J215222.9-283549 & 18.51 & 1.228 & 0.927 &2.59& efl\\
J215342.9-301413 & 19.03 & 1.729 & 1.037 &1.54& efl\\
J215359.0-292108 & 18.75 & 1.160 & 1.036 &3.33& fl\\
J215955.4-292909 & 18.85 & 1.477 & 1.241 &6.24& efl\\
J220003.0-320156 & 19.17 & 2.047 & 1.135 &3.20& fl\\
J220137.0-290743 & 19.36 & 1.266 & 0.600 &3.81& fl\\
J220208.5-292422 & 18.96 & 1.522 & 1.490 &2.98& cfgl\\
J220214.0-293039 & 18.45 & 2.259 & 1.219 &3.39& fl\\
J220650.0-315405 & 19.67 & 2.990 & 2.526 &3.84$^4$& acgk\\
J220655.3-313621 & 19.31 & 1.550 & 0.754 &4.60& efl\\
J220738.4-291303 & 19.57 & 2.688 & 2.666 &1.58$^4$& abcik\\
J221155.2-272427 & 18.80 & 2.209 & 1.390 &2.76& fl\\
J221445.9-312130 & 18.88 & 2.190 & 1.937 &2.24$^5$& cgk\\
J221546.4-273441 & 19.14 & 1.967 & 0.785 &2.86& efl\\
J222849.4-304735 & 19.00 & 1.948 & 1.094 &3.86& fl\\
J223309.9-310617 & 18.64 & 1.702 & 1.146 &2.58& efl\\
\hline  
\end{tabular}
\end{table}
\setcounter{table}{0}
\begin{table}
\caption{Low ionisation 2QZ absorption systems}
\begin{tabular}{lccccl}
\hline  
2QZ QSO ID & $b_J$ & $z_{QSO}$ & $z_{abs}$ &$W_0^2$& Ion$^1$  \\
\hline  
J224009.4-311420 & 19.23 & 1.861 & 1.450 &2.04$^6$& fl\\
J225915.2-285458 & 19.07 & 1.986 & 1.405 &4.57& fl\\
J230214.7-312139 & 19.71 & 1.699 & 0.955 &1.98& fl\\
J230829.8-285651 & 19.06 & 1.291 & 0.726 &3.94& efl\\
J230915.3-273509 & 19.20 & 2.823 & 1.060 &4.66& fl\\
J231227.4-311814 & 18.70 & 2.743 & 1.555 &2.95& fl\\
J231412.7-283645 & 19.31 & 2.047 & 1.920 &3.18$^5$& bcghikl\\
J231459.5-291146 & 18.68 & 1.795 & 1.402 &3.12& fl\\
J231933.2-292306 & 19.55 & 2.013 & 1.846 &4.18$^5$& bcgil\\
J232023.2-301506 & 19.60 & 1.149 & 1.078 &3.50& fl\\
J232027.1-284011 & 19.10 & 1.300 & 1.304 &2.84& efl\\
J232330.4-292123 & 19.47 & 1.547 & 0.811 &3.70& fl\\
J232700.2-302636 & 18.97 & 1.921 & 1.476 &6.41& cefgil\\
&&& 0.673 &3.60& fl\\
J232914.9-301339 & 19.64 & 1.494 & 1.294 &2.78& fl\\
J232942.3-302348 & 19.35 & 1.829 & 1.581 &6.99& cfl\\
J233940.1-312036 & 19.16 & 2.611 & 1.444 &2.34& fl\\
J234321.6-304036 & 18.90 & 1.956 & 1.052 &2.87& fl\\
&&& 1.929 &0.70$^5$& bcdgil\\
J234400.8-293224 & 18.26 & 1.514 & 0.851 &3.13& fl\\
J234402.4-303601 & 19.62 & 0.844 & 0.852 &3.75& efl\\
J234405.7-295533 & 18.55 & 1.705 & 1.359 &2.88& fl\\
J234527.5-311843 & 18.30 & 2.065 & 0.828 &6.67& fl\\
J234550.4-313612 & 18.89 & 1.649 & 1.138 &1.95& efl\\
J234753.0-304508 & 18.56 & 1.659 & 1.421 &2.49& fl\\
J235714.9-273659 & 18.92 & 1.732 & 0.814 &3.52& fl\\
J235722.1-303513 & 19.36 & 1.910 & 1.309 &3.65& fl\\
J000534.0-290308 & 18.90 & 2.353 & 1.168 &4.83& efl\\
&&& 2.226 &2.59$^4$& abcgik\\
J000811.6-310508 & 19.05 & 1.683 & 0.715 &2.55& fl\\
J001123.8-292500 & 18.50 & 1.275 & 0.605 &6.82& efl\\
J001233.1-292718 & 19.07 & 1.565 & 0.913 &2.83& efl\\
J002832.3-271917 & 18.52 & 1.622 & 0.753 &1.83& fl\\
J003142.9-292434 & 18.76 & 1.586 & 0.930 &5.39& efl\\
J003533.7-291246 & 20.00 & 1.492 & 1.457 &3.78& efl\\
J003843.9-301511 & 18.58 & 1.319 & 0.979 &2.93& fl\\
J004406.3-302640 & 19.68 & 2.203 & 1.042 &3.10& fl\\
J005628.5-290104 & 18.60 & 1.809 & 1.409 &3.63& efl\\
J011102.0-284307 & 18.59 & 1.479 & 1.156 &3.24& fl\\
J011720.9-295813 & 19.36 & 1.646 & 0.793 &2.53& fl\\
J012012.8-301106 & 18.96 & 1.195 & 0.684 &4.01& fl\\
J012315.6-293615 & 18.66 & 1.423 & 1.113 &2.30& efl\\
J012526.7-313341 & 19.07 & 2.720 & 2.178 &5.03$^4$& abcdgijkl\\
J013032.6-285017 & 19.32 & 1.670 & 1.516 &3.37& fl\\
J013356.8-292223 & 20.09 & 2.222 & 0.838 &4.64& efl\\
J013659.8-294727 & 18.43 & 1.319 & 1.295 &3.01& fl\\
J014729.4-272915 & 19.29 & 1.697 & 0.811 &3.92& fl\\
J014844.9-302817 & 18.39 & 1.109 & 0.867 &1.75& fl\\
J015550.0-283833 & 20.09 & 0.946 & 0.677 &2.62& fl\\
J015553.8-302650 & 19.49 & 1.512 & 1.316 &3.18& fl\\
J015647.9-283143 & 19.63 & 0.919 & 0.884 &3.14& fl\\
J015929.7-310619 & 18.98 & 1.275 & 1.079 &1.56& fl\\
J021134.8-293751 & 18.97 & 0.786 & 0.616 &3.45& efl\\
J021826.9-292121 & 19.20 & 2.469 & 1.205 &4.45& efl\\
J022215.6-273231 & 19.20 & 1.724 & 0.611 &2.55 & efl\\
J022620.4-285751 & 18.43 & 2.171 & 1.022 &9.03& efl\\
J023212.9-291450 & 19.67 & 1.835 & 1.212 &4.63& fl\\
&&& 1.287 &3.06& fl\\
J024824.4-310944 & 18.59 & 1.399 & 0.789 &4.91& efl\\
&&& 1.371 &1.37& efl\\
J025259.6-321125 & 18.78 & 1.954 & 1.735 &3.94$^3$& cfghkl\\
J025607.9-311732 & 18.78 & 1.255 & 0.903 &3.78& efl\\
&&& 0.975 &4.62& efl\\
\hline  
\end{tabular}
\end{table}
\setcounter{table}{0}
\begin{table}
\caption{Low ionisation 2QZ absorption systems}
\begin{tabular}{lccccl}
\hline  
2QZ QSO ID & $b_J$ & $z_{QSO}$ & $z_{abs}$ &$W_0^2$& Ion$^1$  \\
\hline  
J025919.2-321650 & 19.34 & 1.557 & 1.356 &3.15& fl\\
J030249.6-321600 & 18.27 & 0.898 & 0.821 &4.54& fl\\
J030324.3-300734 & 18.69 & 1.713 & 1.190 &2.95& fl\\
J030647.6-302021 & 19.03 & 0.806 & 0.745 &4.21& fl\\
J030711.4-303935 & 19.09 & 1.181 & 0.966 &2.81& efl\\
&&& 1.108 &2.79& efl\\
J030718.5-302517 & 19.48 & 0.992 & 0.711 &4.95& efl\\
J030944.7-285513 & 19.35 & 2.117 & 0.931 &3.39& fl\\
J031255.0-281020 & 19.21 & 0.954 & 0.953 &2.06& fl\\
J031309.2-280807 & 19.15 & 1.435 & 0.950 &1.78& fl\\
J031426.9-301133 & 18.41 & 2.071 & 1.128 &6.08& efl\\
&&& 1.631 &1.20$^3$& fl\\
\hline  
\end{tabular}

$^1$ Ion identifications:- \\
a) H$\;$I b) C$\;$II c) C$\;$IV d) O$\;$I e) Mg$\;$I f) Mg$\;$II \\
g) Al$\;$II 
h) Al$\;$III i) Si$\;$II j) Si$\;$III k) Si$\;$IV l) Fe$\;$II \\
$^2$ Rest equivalent width of Mg$\;${\small II} $\lambda\lambda 2796,2803$ doublet (\AA) unless stated explicitly below:\\
$^3$ Rest equivalent width of Fe$\;${\small II} $\lambda 2600$ (\AA) (Mg$\;${\small II} blended with sky line)\\
$^4$ Rest equivalent width of H$\:${\small I} $\lambda 1215$ (\AA)\\
$^5$ Rest equivalent width of C$\:${\small IV} $\lambda\lambda 1548,1550$ doublet (\AA)\\
$^6$ Rest equivalent width of Mg$\:${\small II} $\lambda 2796$ (\AA) only\\
\end{table}

\section{QSO Absorbers and Large Scale Structure} 
 The largest heavy element absorption systems, DLAs, are 
generally believed to be the progenitors of
present day spiral galaxies (Wolfe 1995). They dominate the mass of
neutral gas at redshift $z\simeq3$, a mass comparable with that of the
stars in present day spiral disks, suggesting that DLAs are the source
of most of material available for star formation at high redshift
(Lanzetta, Wolfe \& Turnshek 1995). 21cm emission surveys of the local universe
have found that the contribution from intergalactic gas to the
H$\:${\small I} cross-section is negligible, whereas spirals
contribute around 90 per cent (Rao \& Briggs 1993), adding further
weight to this hypothesis.
HST imaging of QSO fields has been carried out in an attempt to
determine the morphology and size of DLA absorbers. A wide range of
luminosities and morphologies were seen, a result inconsistent with
the standard H$\:${\small I} disk paradigm (Le Brun et al. 1997; Rao
\& Turnshek 1998). Pettini et al. (1999) showed
further that the metallicity of DLAs does not increase towards solar
values with decreasing redshift, as expected by the large spiral
progenitor model, but instead little evolution is seen, with
metallicities staying sub-solar. It is possible that dust in the most
metal rich galaxies obscures light from any background QSO making
these objects too faint to see (Pei \& Fall 1995). If so, this dust
bias could help explain why the observed DLAs appear metal poor in
comparison to present day galaxies. Another possible explanation is
that the absorption cross-section may be dominated by diffuse objects
with relatively low star-formation rates, whereas rapid star formation
occurs in the most compact galaxies (Mo, Mao \& White 1999). This evidence
all points towards a mixed population of galaxies, with a large
 fraction of low-mass, or perhaps low surface brightness galaxies,
being responsible for the DLA absorbers, rather than the simpler
spiral progenitor model (Vladilo 1999).

QSO imaging experiments at redshifts out to $z\sim$2 suggest that the
clustering environment of UVX QSOs is similar to optically selected
galaxies (Smith, Boyle \& Maddox 1995; Croom \& Shanks 1999). By measuring the two-point auto-correlation function of the 2QZ QSOs, Croom et al. (2001b) compared the clustering of QSOs at redshift $\bar{z}\sim1.4$ with the clustering of normal galaxies locally ($\bar{z}\sim0.05$) and concluded that the clustering properties are very similar. It would be preferable, however, to compare directly the clustering of QSOs and galaxies in the same redshift intervals. Whilst determining accurate redshifts for a large sample of high-redshift optically selected galaxies is challenging, absorption line
analysis offers a complementary approach to constraining the
QSO-galaxy bias. 

To examine how the absorption systems cluster relative to the 2QZ QSOs, we calculate the two-point absorber-QSO cross-correlation function, $\xi_{AQ}(s)$, in an $\Omega_{\rm m}$=0.3, $\Omega_{\Lambda}$=0.7 cosmology. To estimate the effective volume of each bin, a catalogue of unclustered, random points that have the same radial selection function and angular mask, and approximately twenty times the mean density as the 2QZ QSOs is constructed, taking into account Galactic extinction (Schlegel, Finkbeiner \& Davis 1998). The random catalogue number density is normalised to the number of observed QSOs on each UKST plate to correct for possible small residual calibration errors in the relative magnitude zero points of the UKST plates (see Croom et al. (2001b) for further details).  We use the estimator 
$$\xi_{AQ}(s) = \frac{D_{Abs} D_{QSO}(s)\; \bar{n}_R}{D_{Abs} R_{QSO}(s)\; \bar{n}_{QSO}},$$
where $D_{abs} D_{QSO}(s)$ is the number of absorber-QSO pairs at a given separation in redshift space, $s$, and $D_{abs} R_{QSO}(s)$ is the number of absorber-random pairs. As we are determining the average clustering of QSOs about each absorber, the spatial distribution of the absorbers themselves is not important (this distribution is not straightforward as the absorbers are selected along multiple lines-of-sight through the volume). The errors are calculated using the Poisson estimate, which is accurate at small scales, $<50\;h^{-1}$Mpc, because each pair is independent. All absorbers that lie within $\delta z =0.05$ of the emission redshift of the QSO are excluded from the analysis to ensure that we are only considering truly intervening absorbers. In fact, all of the absorber-QSO pairs within $100\;h^{-1}$Mpc separation are due to an absorber in a different line-of-sight to the QSO.

The absorber-QSO cross-correlation function, $\xi_{AQ}(s)$, is shown in Fig.~2. As there are relatively few absorbers in the analysis, the errors are still quite large and clustering is only detected at the 1.5$\sigma$ level. The amplitude of the clustering appears to agree remarkably well with the QSO auto-correlation measurement (Croom et al. 2001b); further evidence that QSOs are clustered like normal galaxies at these high redshifts. By including absorption systems detected in low S/N spectra, and extending this analysis to the full 2QZ dataset when the survey is complete it will soon be possible to measure the cross-correlation function to much higher precision. 

\begin{figure}
\vspace{-2cm}
\centerline{\hbox{\psfig{figure=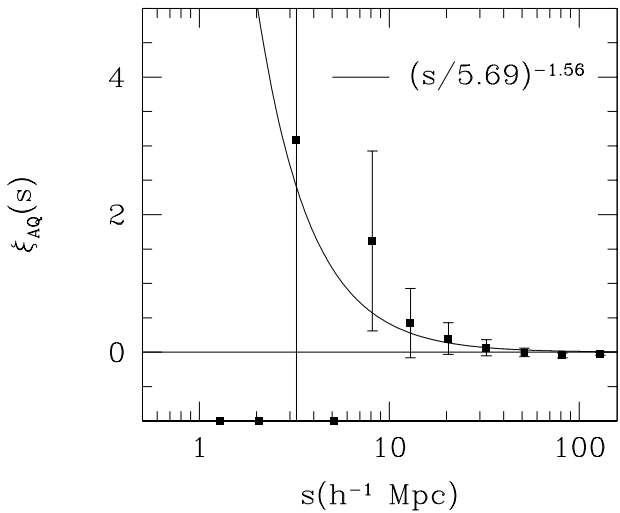,width=9cm}}}
\caption{The two-point cross-correlation function between the QSO absorbers and 2QZ QSOs, $\xi_{AQ}(s)$, calculated in an $\Omega_{\rm m}$=0.3, $\Omega_{\Lambda}$=0.7 cosmology. Overlaid is the best-fitting power law to the QSO auto-correlation function measured by Croom et al. (2001b)
}
\label{fig2}
\end{figure}

\section{QSO colour and dust implications}

The average colour of the QSOs in this sample is $(\overline{B_J-R})_{ABS}=0.546$. To test whether the QSO spectra have been reddened by dust in the foreground absorption systems, we compare this colour to the average colour of QSOs in the 2QZ catalogue. As QSO colour is redshift dependent, we draw random samples from the full catalogue with the same redshift distribution as the QSOs in this absorption line sample to produce a fair comparison. The resultant expected QSO colour for a random sample of 2QZ QSOs of this size and redshift distribution is  $(\overline{B_J-R})_{QSO}=0.493\pm0.026$. Hence we have a $2\sigma$ detection of reddening in the absorption system QSOs. The spectra are reddened by an average of $E(B_J-R)=0.053$, corresponding to a visual extinction $A(V)=0.13$, assuming an $R_V=3.1$ extinction curve (Schlegel, Finkbeiner, \& Davis 1998).

Adopting the relationship between interstellar extinction and neutral hydrogen gas column density from Bohlin, Savage \& Drake (1978), 
$$<N(\mathrm{HI}+\mathrm{H}_2)/E(B-V)>=5.8\times 10^{21}\,atoms\,cm^{-2}\,mag^{-1},$$
 we obtain an average neutral hydrogen column of $N(\mathrm{HI})\sim 2.3\times 10^{20}$\,atoms\,cm$^{-2}$ for our absorbers. This is approximately the minimum column density of a DLA, and is probably considerably smaller than the average neutral hydrogen column of our sample. The reason for this is that the dust-to-gas ratio in high redshift DLAs is typically much lower than in the local interstellar medium. Using zinc and chromium abundance measurements, Pettini et al. (1997) found that the average dust-to-gas ratio in DLAs at $\bar{z}\sim2.5$ is $\sim 1/30$ that in the Milky Way. Our sample of absorbers is at a lower redshift and, assuming an average neutral hydrogen column of $N(\mathrm{HI})\sim 10^{21}$\,atoms\,cm$^{-2}$, the observed extinction appears consistent with a dust-to-gas ratio of about a quarter that of the Milky Way.  It is possible that selection effects could bias our sample towards higher metallicity or column density. Further observations could be used to determine these quantities, and hence allow a more reliable measurement of the dust in these systems.

\section*{Acknowledgements}

\noindent The 2QZ is based on observations made with the Anglo-Australian Telescope and the UK Schmidt Telescope, and we would like to thank our colleagues on the 2dF galaxy redshift survey team and all the staff at the AAT that have helped to make this survey possible. NL acknowledges the receipt of a PPARC studentship.


\begin{thebibliography}{}
\bibitem[\protect\citename{Bailey \& Glazebrook\ }
1999]{bai99} Bailey, J.A., Glazebrook, K., 1999, 2dF User Manual, Anglo-Australian Observatory
\bibitem[\protect\citename{Bohlin, Savage \& Drake\ }
1978]{boh78} Bohlin, R.C., Savage, B.D., Drake, J.F., 1978, ApJ, 224, 132
\bibitem[\protect\citename{Croom \& Shanks\ }
1999]{cs99} Croom, S.M., Shanks, T., 1999, MNRAS, 303, 411
\bibitem[\protect\citename{Croom et al.\ }
2001a]{cro01a} Croom, S.M., Smith, R.J., Boyle, B.J., Shanks, T., Loaring, N.S., Miller, L., Lewis, I.J., 2001a, MNRAS, 322, L29
\bibitem[\protect\citename{Croom et al.\ }
2001b]{cro01b} Croom, S.M., Shanks, T., Boyle, B.J., Smith, R.J., Miller, L., Loaring, N.S., Hoyle, F., 2001b, MNRAS, in press.
\bibitem[\protect\citename{Fan et al.\ }
1999]{fan99} Fan, X., et al. 1999, AJ, 118, 1
\bibitem[\protect\citename{Lanzetta et al.\ }
1995]{lan95} Lanzetta, K.M., Wolfe, A.M., Turnshek, D.A., 1995, ApJS, 440, 435
\bibitem[]{l7} Le Brun, V., Bergeron, J., Boiss{\'e}, P., Deharveng, J.M. 1997, A\&A, 321, 733
\bibitem[\protect\citename{Lewis, Glazebrook \& Taylor\ }
1998]{lew98} Lewis, I.J., Glazebrook, K., Taylor, K., 1998, Proc. SPIE, 3355, 828
\bibitem[]{m12} Mo, H.J., Mao, S., White, S.D.M. 1999, MNRAS, 304, 175
\bibitem[]{p2} Pei, Y.C., Fall, S.M. 1995, ApJ, 454, 69
\bibitem[\protect\citename{Pettini et al.\ }
1997]{pet97} Pettini, M., King, D.L., Smith, L.J., Hunstead, R.W., 1997, ApJ, 478, 536
\bibitem[]{p14} Pettini, M., Ellison, S.L., Steidel, C.C., Bowen, D.V. 1999, ApJ, 510, 576
\bibitem[]{r1} Rao, S.M., Briggs, F. 1993, ApJ, 419, 515
\bibitem[]{r2} Rao, S.M., Turnshek, D.A. 1998, ApJ, 500, L115
\bibitem[\protect\citename{Schlegel, Finkbeiner, \& Davis\ }
1998]{sch98} Schlegel, D.J., Finkbeiner, D.P., Davis, M., 1998, ApJ, 500, 525
\bibitem[\protect\citename{Smith et al.\ }
1995]{smi95} Smith, R.J., Boyle, B.J., Maddox, S.J., 1995, MNRAS, 277, 270
\bibitem[\protect\citename{Smith et al.\ }
2001]{smi01} Smith, R.J., Croom, S.M., Boyle, B.J., Shanks, T., Miller, L., Loaring, N.S., 2001, MNRAS, submitted.
\bibitem[]{v4a} Vladilo, G. 1999, to appear in {\it The Evolution of Galaxies on Cosmological Timescales}, eds. Mahoney, T.J., Beckman, J.E., ASP Conf. Ser. 
\bibitem[\protect\citename{Wolfe\ }
1995]{wol95}  Wolfe, A.M., 1995, in {\it QSO Absorption Lines}, ed. Meylan, G., Springer-Verlag, 13
\end{thebibliography}
\end{document}